\documentclass[prl,twocolumn,amsmath,amssymb]{revtex4-1} 
\usepackage{appendix}
\usepackage{graphicx}
\usepackage{epsfig} 
\usepackage{caption}
\usepackage{subfloat}
\usepackage[dvipsnames]{xcolor}
\usepackage[T1]{fontenc}
\usepackage{wrapfig}
\usepackage[english]{babel}
\usepackage{mathrsfs} 
\usepackage{amsmath}
\usepackage{amsfonts}
\usepackage{sidecap}
\usepackage{bbm}
\usepackage[multidot]{grffile}
\usepackage{subfig}
\usepackage{multirow}
\usepackage{url}
\usepackage[english]{babel}
\usepackage{graphicx}
\usepackage{dcolumn}
\usepackage{bm}
\usepackage{times}
\usepackage{float}
\usepackage{physics}
\usepackage[small,compact]{titlesec}
\usepackage{hyperref}   
\usepackage[utf8]{inputenc}
\usepackage[T1]{fontenc}
\usepackage{hyperref}
\usepackage{xr}
\newcommand{\be}{\begin{equation}}
\newcommand{\ee}{\end{equation}}

\makeatletter
\renewcommand*\env@matrix[1][*\c@MaxMatrixCols c]{%
  \hskip -\arraycolsep
  \let\@ifnextchar\new@ifnextchar
  \array{#1}}
\makeatother

\begin{document}

\title{Self-assembly of correlated Kondo lattices: The Mott to Kondo transition in diluted superlattices}

\author{Hovan Lee}
\author{Evgeny Plekhanov}
\author{David Blackbourn}
\author{Swagata Acharya}
\author{Cedric Weber}
\affiliation{King's College London, Theory and Simulation of Condensed Matter, The Strand, WC2R 2LS London, UK}
\maketitle


\textbf{In the field of condensed matter, the quest to obtain an experimental realization of a Kondo lattice has generated a tremendous effort of the community, from both standpoints of experiments and theory. The pursuit of obtaining independent magnetic moments, via charge localization through Coulomb interactions, is paramount for applications in nanotechnology. In particular, systems with simultaneous charge and spin degrees of freedom can manifest both Kondo spin quenching and Mott-Hubbard charge localization. A unified experimental framework illuminating the pathway between the two phenomena is of physical and technological interest, and is (as of yet) hardly observed in real condensed matter systems. 
Recent developments in the ability to control densities and temperatures of strongly correlated Fermionic impurities on surfaces and substrates has opened up a new paradigm of possibilities for this pathway. In particular, a milestone was recently surpassed through the observation of self-assembled superlattices of f band adatoms on metallic surfaces, such as the deposition of Ce on Ag(111). Such lattices have introduced a mechanism of diluted correlated lattices where the interaction between Kondo and Mott physics can be methodically studied. However, it remains difficult to control the adatom distances and substrate densities in these systems, and the interplay between Kondo physics and charge localization remains elusive. In this work, we systematically investigate the phase diagram of superlattice structures of heavy f elements deposited on metallic substrates, and assess the required conditions to obtain Kondo lattices in superlattices. We unveil a unique pathway between Kondo quenching and Mott localization, and identify
a non-trivial charge density wave phase emerging from the competition of charge localization and Kondo physics.}

Deposition of f elements upon metallic substrates have generated a tremendous interest in the scientific community due to the formation of self-assembled f-element superlattices. In particular, this has opened avenues towards experimental realization of Kondo lattices. 
A number of self-assembled structures have been discovered, however no clear theoretical description has provided a classification of the Mott and Kondo physics interplay of these experiments as of yet. In this work we provide  a general phase diagram of correlated super-structures as a function of correlation strength, substrate electronic density, and adatom inter-distances. Through the use of state-of-the-art dynamical mean-field theory calculations at finite temperature, we have identified a clear regime of parameters where Kondo lattices can be realized. We also report a sharp transition between Mott type physics (for short adatom inter-distances) and Kondo physics. We also report the stabilization of a charge density wave competing with Kondo physics at large inter-adatom distances. Finally, for half-filled f-occupation the ionic potential of the adatom induces a non trivial bound state between the f band electrons and substrate electrons. 


Correlated adatoms expand upon the plethora of remarkable properties already existent in two-dimensional (2D) metallic surfaces. Stemming from the competition of localization and itinerant characters, such properties include: hybridization, superlattice self formation\cite{blackbourne,self1,self2} and long range interactions\cite{lr1,lr2,lr3,lr4}. Moreover, due to the comparatively low coordination between adatoms and substrate, single atom magnets\cite{mag1,mag2,mag3} and transistors\cite{transistor} were discovered. These phenomena have attracted a wide academic interest, but also offer pathways towards industrial applications, for example as  candidates for future atomic scale memory storage, as employed by IBM\cite{ibm}.

\begin{figure}
\vspace{0.6cm}
\includegraphics[width=1.5\columnwidth, trim= 10.0cm 14cm .0cm 13cm]{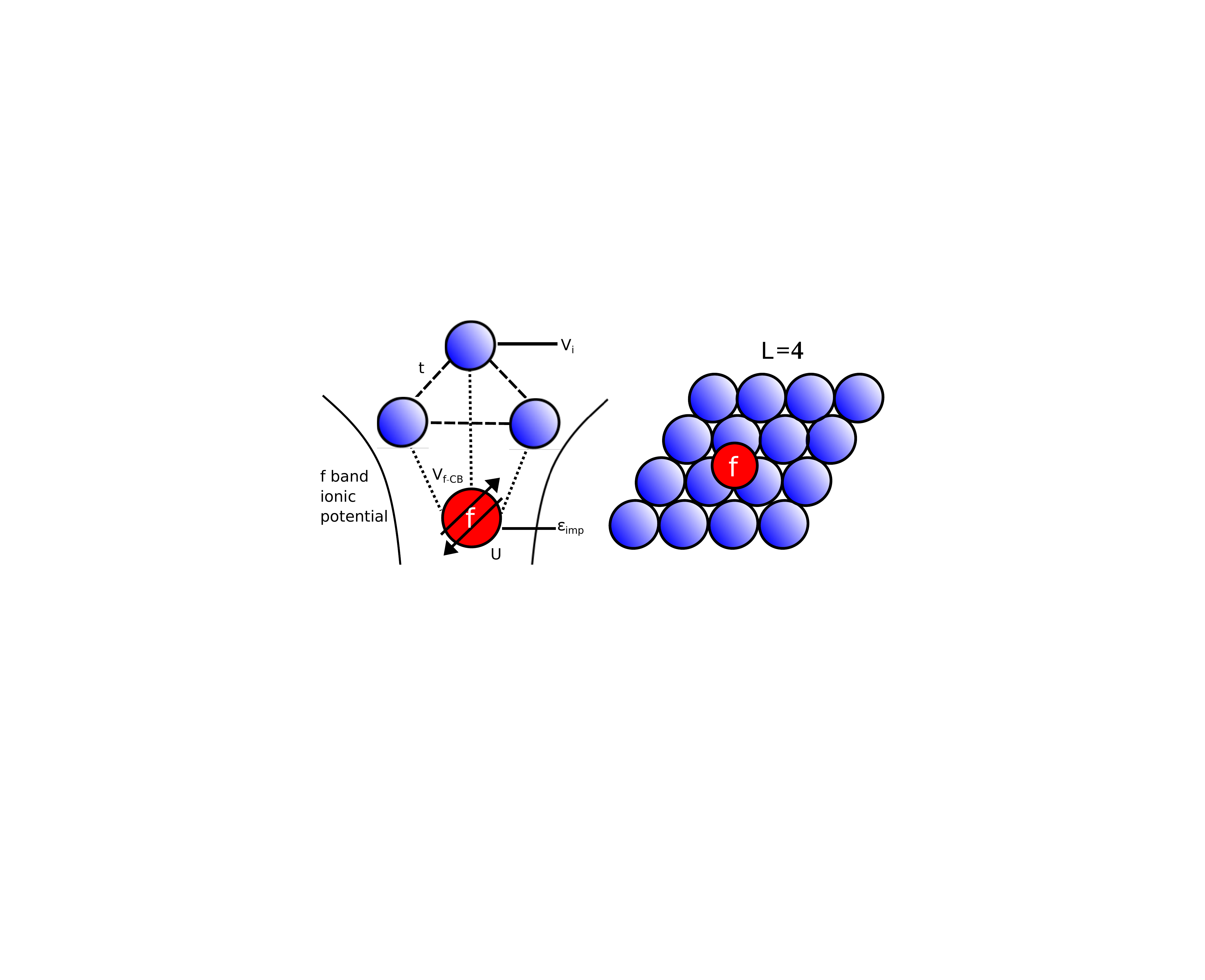}
\caption{(colour online) Cartoon picture of the DMFT setup. The model includes the conduction band of the substrate (depicted by the blue circles) and the concomitant Hamiltonian parameters. The correlated adatom (red circle)
sits in between three sites of the substrate (blue circles). Inter-adatom distance $L$ is parameterized by the number of substrate sites between each adatom.}
\label{scheme}
\end{figure}

While the limits of Mott\cite{mott_exp,mott_thy_exp,mott_thy_exp_2} and Kondo\cite{kondo_thy,kondo_thy_2,kondo_thy_3} superlattice behaviour are well documented, little is known about the transition between the densely packed (Mott) and the diluted (Kondo) limits of  superlattice systems. Furthermore, although physical cases of $3d$ transition adatom\citep{3d1,3d2,3d3,3d4} and rare-earth adatom\cite{re1,re2,re3} self organization have been reported and justified with tight-binding models and kinetic Monte Carlo simulations\cite{cedric2004,sl1,sl2,sl3,sl4}, the quantum mechanism responsible for the formation of the superlattice, and the  adatom-substrate requirements for self formation are so far elusive. 

\begin{figure*}[th]
\includegraphics[width = 1.075\textwidth, trim= 3cm 1cm 0cm 1.25cm]{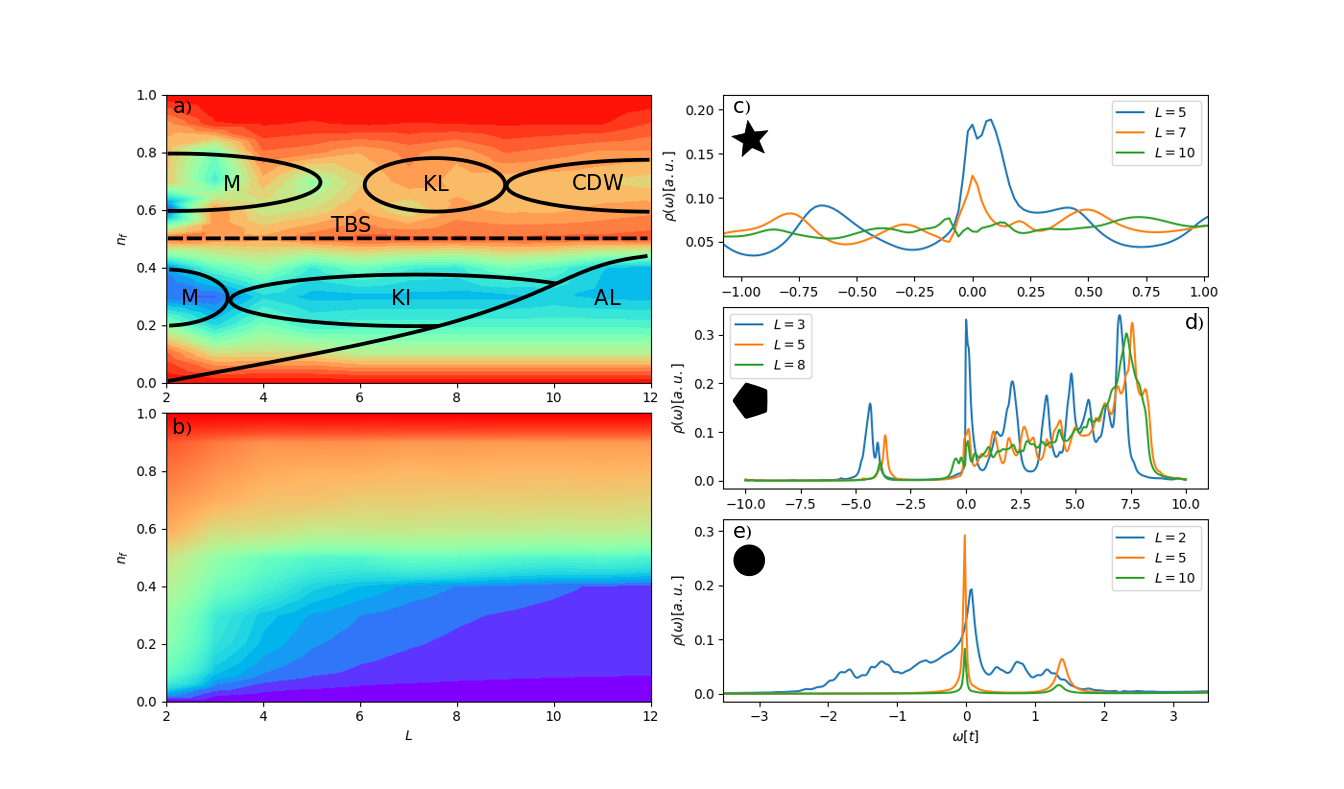}
\caption{(colour online) a): colour map of the quasi-particle weight $Z$ for $U=6$ and different lattice sizes $L$. $Z$ is a measure of the strength of the electronic correlations induced by $U$, where the colour map indicates $Z=0$ (blue) and $Z=1$ (red). We identified three regions of interest in the phase diagram: for $n_f=0.3$, $n_f=0.5$ and $n_f=0.7$. Each region is analyzed and split into phases (M - Mott, KI - Kondo impurity, AL - atomic limit, TBS - triplet bound state, KL - Kondo lattice and CDW - charge density wave). We report further in the text observables that characterize these phases. b): colour map of the substrate conduction band (CB) electronic density different substrate sizes $L$ (blue for $n_{CB}=0$ and red for $n_{CB}=1$). The substrate's electronic states are empty in the lower right corner of the phase diagram, which indicates that in this region the adatom is decoupled from the substrate. Right column: Total system density of states, across different regions of $n_f$. Each set of data is renormalized with the supercell size and shifted with respect to the chemical potential. This was done to allow for comparison between datasets. From top to bottom, the adatom electron densities are $0.7$, $0.5$, $0.3$. The star, pentagon and circle symbols indicate points on the phase diagram (see fig.S1)}
\label{atlas}
\end{figure*}

To paint a picture of the phenomena observed in this work, an overview of impurity lattice physics is due: a single Kondo adatom within a material does not conserve momentum when scattering electrons. However, in this work we discuss lattices whose unit cell contains Kondo impurities: Kondo lattices. Due to translational symmetries, Kondo scattering within such lattices conserve momentum, leading to coherent scattering and a decrease in resistivity below the Kondo temperature\cite{kondolattice,kondolattice2}. Interpreting requirements for constructing Kondo lattices will be essential in further understanding of correlated low temperature research. Friedel oscillations\cite{friedel} are another form of impurity-impurity communication. They describe electronic screening of an impurity and charge distribution on lattice sites. More specific to this system is a theory presented by Hyldgaard and Persson\cite{handp} on how Friedel oscillations from multiple impurities interact. Lastly, RKKY interactions\cite{rkky1,rkky2,rkky3} provide a mechanism for spin-spin coupling between f and/or d band electrons. 

2D systems promote surface electron states. These states are formed with lower electron densities in comparison to the bulk electron states observed in three-dimensional systems. Therefore, electrons become localized under the effects of impurities, and the system becomes devoid of conduction band electrons. Since there is no scattering of conduction band electrons, this leads to an atomic limit behaviour at the lattice impurities. Moreover, substrate electrons are further localized by the impurity induced potential. Under these circumstances, a spin-spin interaction can occur between the impurities electrons and electrons localized near the impurities, forming singlet/triplet bound states. Lastly, due to the periodic structure of both the lattice and the impurity ionic potential, localization can occur around both the adatoms and on the substrate. Localization on substrate sites might be expected when inter-adatom distances become sufficiently large, and the conduction band of the substrate sufficiently high, permitting localization effects and realizing a charge density wave concomitant with a pseudo-gap formation. These effects have so far not been investigated, due to the complexity of the problem, which involves dealing with large unit-cells and the presence of strong correlations in the f-shell of the adatoms. Typically, such systems are beyond the reach of density functional theory, and even more so, from GW and its extensions. 

In this work, we use a combination of Dynamical Mean Field Theory (DMFT) and tight binding calculations for a triangular lattice substrate. We present our investigations into the generic properties of single orbital correlated adatom structure upon a triangular metallic substrate. This model is a canonical model for the description of a 4f1 orbital hybridized to a Ag(111) substrate.
We systematically investigate the phase diagram upon different density and adatom distances, and outline the different phases obtained. We study the strength of correlations, quantified by the quasi-particle weight factor ($Z$), and examine the electron localization at the adatom and its impact on the correlated energy bands and Fermi surfaces.

We report a Mott to Kondo transition as adatom spacing varies from a tightly packed solid to a diluted solid. For any given superlattices, large mass renormalization are obtained at low and high density regimes, corresponding
to the limit of a weakly hybridized impurity, and the limit of a screened impurity respectively. These two phases
are separated by the formation of a tightly bound triplet between the adatom and the substrate that exhibits a weak mass renormalization.
In our view, the experimental realization of the Ce/Ag(111) superlattice falls in the latter regime, and our result shed light
on the lack of evidence for Kondo physics as previously observed by spin-polarized STM. Our work predicts that Kondo Lattices are obtained at intermediate adatom distances (5 to 9 substrate lattice sizes, $L$). 

\section{Results \& Discussion}

\subsection{Strength of correlations and emergent quasi-particles in the superlattice}

First, we report the systematic study of the superlattice correlation strength, as function of the substrate electronic density and adatom distances, obtained by paramagnetic DMFT calculations (permitting fluctuating moments and correlations across adatoms, but not for long-range magnetic order). Correlation strength is measured by the adatom quasi-particle weight factor $Z$, obtained with the low energy imaginary component of Matsubara self energy $\Sigma(i\omega_n)$, reported in fig.\ref{atlas}. The results outlined below correspond to Hubbard repulsion $U/t=6$, where $t$ is the tunnelling amplitudes across the substrate, typically $t=0.75$eV for the case of the Ag/Ce superlattice (we discuss results for this Coulomb repulsion thereafter, extended the calculations to
other values of the Coulomb repulsion, see fig.S1 in Supplementary Information). 

We report a phase diagram in terms of the adatom electronic occupation $n_f$ (all occupation refer to the single spin component only, as the calculations are paramagnetic). Although in our calculations we systematically change the chemical potential, we report the phase diagram in terms of the adatom occupancy. The total occupation is indeed a monotonic function of the adatom density at any finite temperature.  We find two dominant regimes where correlation effects are large, at two nearly horizontal minima near the single spin adatom (f band) electron density $n_{f}=0.3$ and $0.7$, separated by a weakly correlated regime at half-filled adatom $n_{f} = 0.5$ across all values of $U$ (see also fig.S1 and fig.S2).

The low $n_{f}$, we obtain hence that the quasi-particle state is localized at the adatom. As $n_{f}$ increases however, $Z$ also increases, reaching a maximum when the adatom  is exactly half-filled (see fig.\ref{ntot}). Although we might naively expect
that the reduction of the mass renormalization is associated with a delocalization of the charge, we observe that the electron
remain tightly bound to the adatom location. In our view, this is associated with the creation of a two-electron tightly bound
triplet state between the adatom and the substrate. Finally, at densities larger than two-electron per unit-cell, we obtain another localization process near $n_{f}=0.7$ with a concomitant increase of the mass renormalization.

We report the electronic density of the substrate in fig.\ref{atlas}.b (blue denotes an empty substrate, and red the band insulating limit). As expected, at low occupation of the adatom, the substrate is empty, and we obtain the limit of an impurity in vacuum (atomic limit, AL). This limit is not physical and provides a lower bound on the adatom occupation for a given superlattice structure. 
Additionally, at small adatom distances $L$, we observe a rapid filling of the substrate upon increasing the adatom occupation, an effect related to the Coulomb repulsion $U$ at the adatom site, that prevents an occupation of more than one electron and induces a charge transfer to the substrate. 

The energy resolved spectral weight $\rho(\omega)$, obtained respectively at adatom density $n_f=0.3$ (see fig.\ref{atlas}.c), at half-filled adatom density $n_f=0.5$ (see fig.\ref{atlas}.d, which also is concomitant with a filling of two electron in total per unit-cell, meaning there is one electron in the substrate), and at large density $n_f=0.7$ (see fig.\ref{atlas}.e). 

\begin{figure}
\includegraphics[width = 1.5\columnwidth, trim = 9.cm 12cm 3cm 13cm]{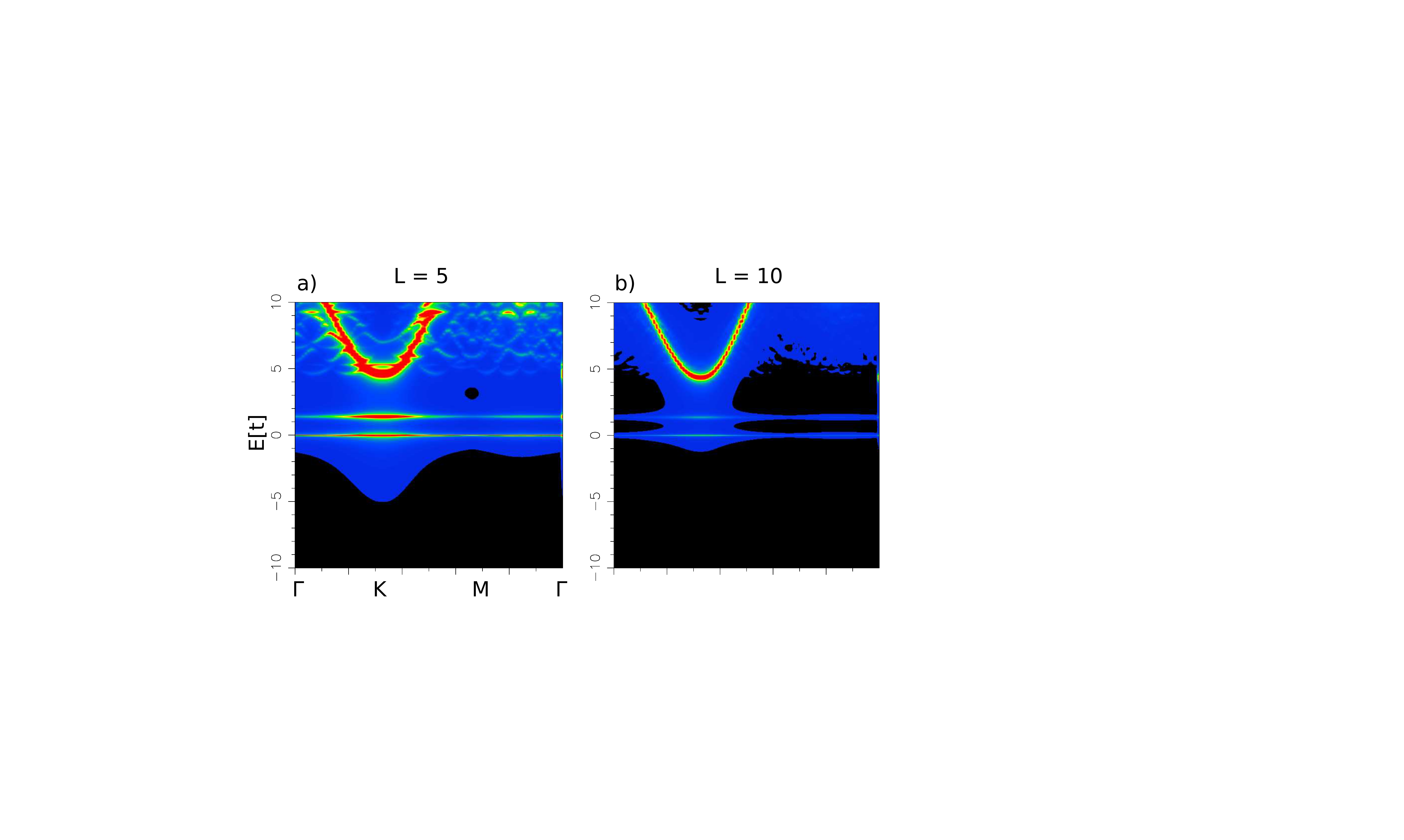}
\caption{(colour online) Spectral weight $A_{k,\omega}$ resolved in momentum and energy for $U=6$ and $n_{f}=0.3$ for a) $L=5$ and b) $L=10$. The adatom f state is strongly hybridized to the conduction band for intermediate lattice sizes ($L=5$), but we observe a dehybridization of the adatom f state and conduction band at larger size $L=10$. For the latter, the adatom is in the atomic limit (AL), as the conduction band remains weakly populated, as shown in fig.\ref{atlas}.e.}
\label{bands}
\end{figure}

The low density regime exhibits (see fig.\ref{atlas}.c) sharp features that corresponds to the localized states at the adatom. We obtain a transition from a broad feature, reminiscent the correlated metal, to a narrow spectral feature at the Fermi level obtained at adatom distances $L>4$. This is fingerprint of a localized electron at the adatom location, which we denote at a Kondo Impurity (KI).
Ultimately, the KI localized quasi-particle states converges to the atomic limit (AL) at large adatom separation. 
In fig.\ref{bands}, we report the spectral function $A(k,\omega)$ resolved in real frequencies $\omega$ and momentum $k$ obtained for the same density. Note that, as the band structure is trivially folded due to the unit-cell construction of the periodic superlattice, we reconstruct the unfolded band structure of the original unit-cell of the substrate, which allows for comparison across different superlattices. For $n=0.3$, we observe that the adatom
are strongly hybridized to the substrate for $L=5$ (fig.\ref{bands}.a), as shown by the fluctuations induced in the substrate bands by the adatom, but dehybridize at larger adatom distance $L=10$ (fig.\ref{bands}.b), a fingerprint that the atomic limit has been reached (AL).

The spectral feature of the weakly correlated region at $n_f=0.5$ (see fig.\ref{atlas}.d) exhibits a large degree of hybridization between the adatom and substrate. We associate this state to a Triplet Bound State (TBS) formation. Finally at larger density $n_f=0.7$ (see fig.\ref{atlas}.e), we observe the emergence of a resonance, that starts to develop at $L=5$, and becomes resonant at $L=7$, and ultimately vanishes at $L=10$. 
We associate this feature to the presence of a screened Kondo singlet (see discussion thereafter). We note that at the same density, we observe at large adatom distances $L>9$ the formation of a charge density wave, where electron on the substrate localize half-way in-between adatoms, with the opening of a pseudo-gap in the spectral density (see fig.S1 and S2). This phase is denoted as CDW in the phase diagram.  

\subsection{The Mott-to-Kondo Crossover in the diluted solid}

To investigate the correlated regions of the phase diagram (fig.\ref{atlas}.a), a polynomial regression was preformed on the adatom self-energy to obtain the electron scattering rate and the Kondo temperature ($T_k$) in the two correlated phases obtained at respectively $n_f=0.3$ and $n_f=0.7$. In both phases, we obtain a significant mass renormalization, concomitant with a large magnetic moment (see fig.S5), a required but not sufficient condition to obtain Kondo physics.


\begin{figure}
\includegraphics[width = 0.85\columnwidth,trim= 1.25cm 0.5cm .85cm .5cm]{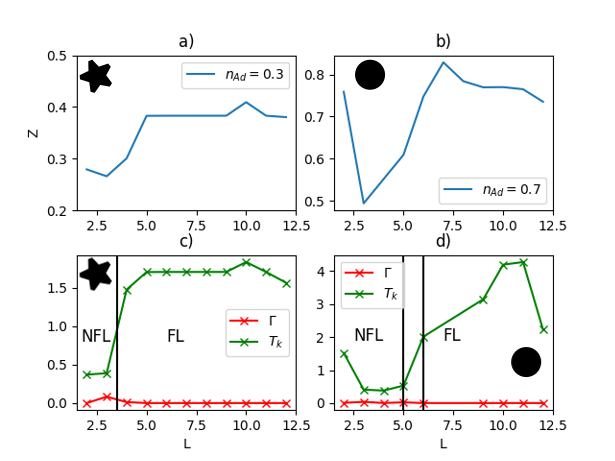}
\caption{(colour online) Quasi-particle weight $Z$ across a range of lattice sizes $L$ for $U=6$ at adatom electron density a) $n_{f}\sim0.3$ and b) $n_{f}\sim0.7$. 
Scattering rate $\Gamma$ and Kondo temperature $T_k$ (as obtained by the formalism detailed in the method section) for c) $n_{f}\sim0.3$ and d) $n_{f}\sim0.7$.
At low density ($n_{f}\sim0.3$), 
we observe a sharp first order transition between non-Fermi liquid NFL and Fermi liquid FL for $L=4$ (see the sharp onset of $T_K$ at $L=4$, concomitant with a reduction of the effective mass $m*$). FL properties are eventually suppressed in the diluted limit ($L>12$), as the atomic limit is obtained with the absence of electronic states in the substrate (see fig.\ref{atlas}.e). For large density ($n_{f}\sim0.7$)
we observe a similar NFL/FL transition at $L=6$.}
\label{gamma}
\end{figure}
The Kondo temperature 
$T_k$, extracted from the low frequency
hybridization (see method section) was used here as a marker of the formation of a Kondo singlet. Note that $T_k$ as extracted in this approximation provides the right trend, but is known to overestimate $T_k$. We obtain (see fig.\ref{gamma}.c and d) that a cross-over occurs for both regions as the adatom distances reaches $L=3-4$ for $n_f=0.3$ and $L=5-6$ for $n_f=0.7$ respectively. For the limit of the Kondo impurity ($n_f=0.3$), the system enters into the Kondo regime abruptly via an abrupt transition (see SI fig.S2) between a non Fermi liquid at short adatom distances to a Fermi liquid at intermediate adatom distances (a linear behavior of the self energy as a function of the Matsubara frequency is obtained in the self energy, a signature of a Fermi liquid, see fig.S2).

For $n_{f}=0.7$, the crossover is continuous (see fig.\ref{gamma}) and, unlike the abrupt change obtained between the NFL to FL regime obtained at $n_{f}=0.3$, an increase in the mass renormalization occurs simultaneously with the large increase of $T_k$ at $n_f=0.7$.

\begin{figure}
  \centering
  \includegraphics[width=.5\textwidth, clip=true, trim = 0.3cm 5.5cm 1cm 1.3cm]{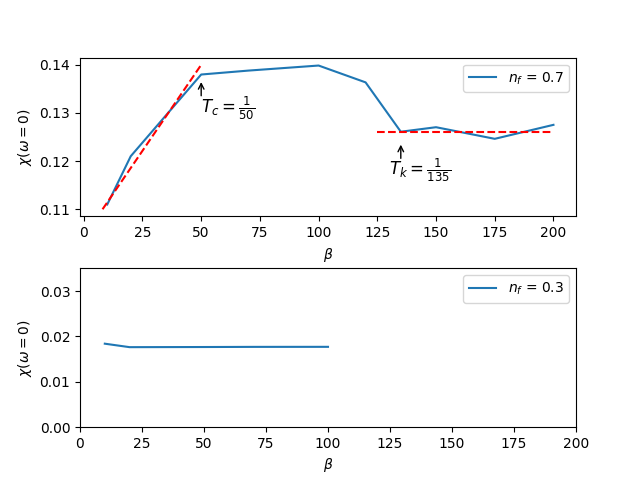}
  \caption{Local magnetic susceptibility $\chi(\omega=0)$ of the adatom for $L=6$ and $n_f=0.7$, as a function of inverse temperature $\beta$. At high temperature ($T>T_c$), the susceptibility is proportional to $1/T$, indicating that the system is above the Curie temperature, whereas at low temperature ($T<T_k$), the susceptibility is temperature independent, showing that the system has entered into the Kondo regime.}
  \label{spin sus}
\end{figure}

The calculation of the Kondo temperature has been validated by the measurement of the static local magnetic susceptibility $\chi(\omega=0)$ of the adatom, obtained from calculations carried out at different temperatures. 
In fig.\ref{spin sus}, we observe that the system manifests true Kondo behaviour in the large density regime $n_f=0.7$, for temperatures lower than $T=0.007 \times t$, which corresponds to $\approx 60^\circ$K \footnote{The tunneling amplitude for the silver substrate is $t=0.75$eV}. We therefore identify the region of the phase diagram $L>5$ and $n_f=0.7$ as a Kondo lattice (KL): a lattice of independent screened magnetic moments. At large temperature $T>0.02 \times t$, we recover the usual Curie Weiss law, which describes a fluctuation moment. 

In the low density limit ($n_f=0.3$), we find that the local susceptibility remains small at all temperatures with weak temperature dependence ($\chi\sim0.02$), as expected for a system which exhibits a large degree of localization due to correlation effects. We also note that a larger Coulomb repulsion $U/t=10$ opens a gap and suppresses the resonance obtained in this regime (see fig.S3.b), gaping in turn the fluctuations associated with the magnetic moment by Mott physics. 
Due to the low occupation of the conduction band, we consider the region $L>3$ and  $n_f=0.3$ as a system of Kondo impurities (KI).

\subsection{Weakly Correlated Triplet Bound State}

\begin{figure}
\includegraphics[width=0.9\columnwidth,trim= 1.0cm 0.5cm 1.0cm 1.3cm]{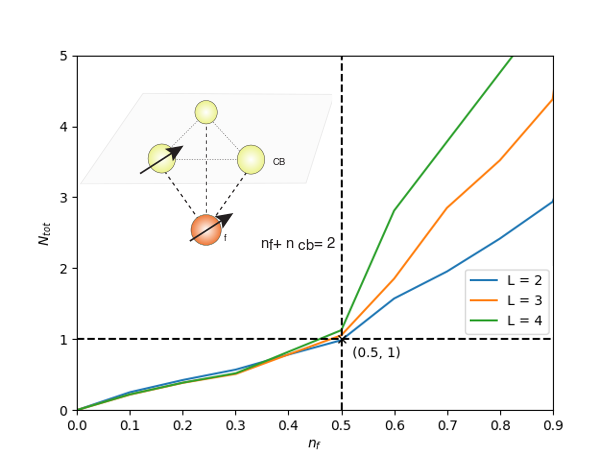}
\caption{(colour online) Number of electrons within the supercell $N_{tot}$ as a function of $n_f$ for $U=6$. Across all lattice size $L$, the graph follows the same linear increase from an empty system to ($n_{f}=0.5,N_{tot}=1$). We conclude that (taking both spins into account) when the adatom is singly occupied, the conduction band must also be singly occupied, allowing singlet/ triplet states between the adatom and substrate to form.}
\label{ntot}
\end{figure}

Finally, we investigated the region of the phase diagram that divides the two correlated regions obtained in the phase diagram (obtained at $n_f=0.3$ and $n_f=0.7$ respectively). At $n_f=0.5$, which corresponds to the case of a half-filled f level, we obtain the weakest mass renormalization and a regime of nearly free electrons (NFE) (see fig.\ref{atlas}.a). In fig.\ref{ntot}, we report the integrated occupation over the supercell (both adatom and substrate). In particular, we find that the half-filled adatom f shell is obtained with a concomitant total charge of 2 electrons, and hence an occupation of a single electron per unit-cell on the substrate. The real space average charge (not shown) also reveals that the electron pair is localized in the unit cell around the adatom. This can be rationalized in a simple toy model, where 2 electrons are forming a bound state between the adatom and the three substrate neighbours (see fig.\ref{ntot}). This toy model can be used as an approximation to capture the physics of the superlattice locally. The solution of the simple model gives a bound triplet ground state for any realistic values of the screened Coulomb interaction $U$ ($U<11eV$). For $U>11eV$, we obtain a singlet ground state. 
It is worth noting that this limit was obtained by renormalizing the hopping parameters between the substrate atoms to recover the bandwidth of the conduction band, as those have been integrated out in this simple picture (without this renormalization, the $U$ separating the singlet from the triplet is larger). We hence conclude that for this region of the phase diagram, we obtain triplet bound electron pairs localized around the adatoms. Note that the triplet can be seen as an uncorrelated state, as the Pauli principle excludes the double occupations. In our view, this phenomena explains the nearly free electron 
regime, separating the correlated phases at low and high $n_f$. 

\subsection{Mott localization driven by adatom interactions mediated by the substrate}

\begin{figure*}
\includegraphics[scale=0.3, trim=11.5cm 11cm 9cm 11cm]{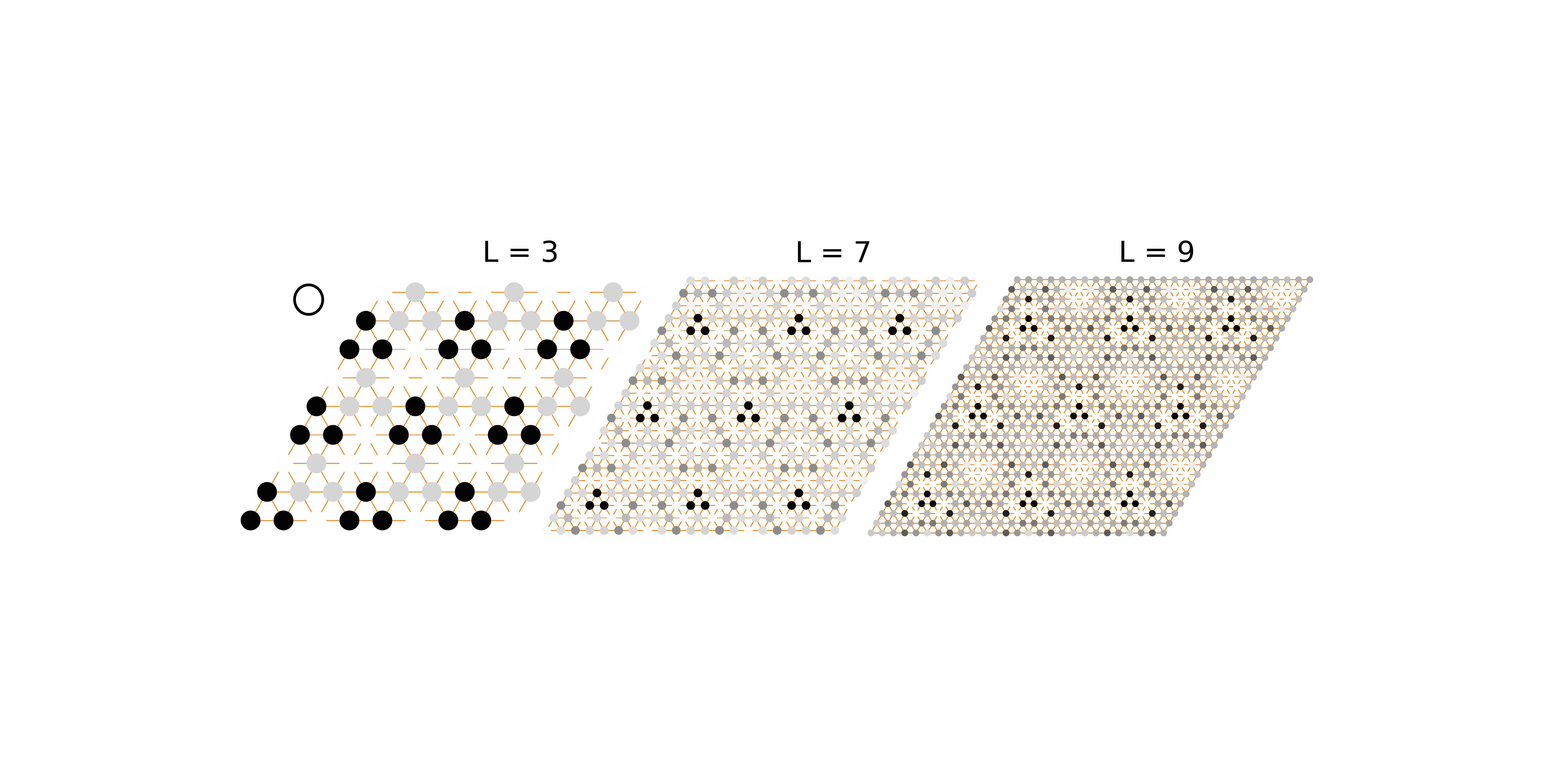}
\caption{Real space electronic density $n_{CB}$ of the substrate conduction band for the strongly coupled regime and $n_{f}\sim 0.7$ for $L=3, 7$ and $9$. Black (white) colors indicate large (small) electronic density.
For $L=3$, we observe localized states around the adatom (the adatom, not shown, is located at the center of the black circles). As $L$ increases, we observe a delocalization, which is concomitant with Kondo physics (see Kondo peak in fig.S3.f and large Kondo temperature in fig.\ref{gamma}). For the limit of diluted solids, corresponding to $L=9$ and larger, we observe the formation of a charge density wave (CDW), where localized states develop in between adatoms. The CDW is concomitant with the formation of a pseudogap (see fig.S3.f).}
\label{real}
\end{figure*}

We also extended our calculations to larger and smaller values of the correlation $U$.
Whilst most results remain identical, we note that for the strong coupling regime ($U/t=10$), the systems undergoes a charge localization for $L=9$ for $n_f=0.7$, followed by the formation of a pseudo-gap and the suppression of the Kondo resonance (see fig.S3.e). The localization is directly observed in the real space electronic density (see fig.\ref{real}.c).
Albeit it is observed in all calculations that electrons localized naturally around the adatoms, due to the attractive ionic potential, the localization induced by the Coulomb repulsion in this particular regime of parameters happens in-between the adatoms. 
In our view, this effect is driven by the two-body potential induced by the interference energies between Friedel oscillations (Hyldgaard-Persson). This two-body adatom interaction, mediated by surface state electrons, is an oscillating function of the adatom distance (similarly to the RKKY interaction), and is parametrized by the electronic mass. In the free electron model, its oscillations are involving distances larger than 11 lattice spacings, and is hence does not play a role in our calculations for $L<11$. Furthermore,
this interaction is also suppressed when the states at the Fermi level are absent (as it is the case in the low density regime $n_f=0.3$ at large $U/t=10$ that opens a gap). 

However, at large density $n_f=0.7$, 
the adatom mediated interaction is 
modified by the large mass renormalisation induced by the local Hubbard U term, and produces an attractive potential in-between the adatoms.

We emphasize that this CDW is fundamentally different from Friedel oscillations (FO), where the amplitude decreases as distance from impurity increases. Secondly, as discovered by Chatterjee et. al.\cite{fo} the amplitude of FO decreases as $U$ increases, in contradiction to our observation in the Kondo superlattice. 
 Furthermore, the charge localization, in our measurements, is obtained in the large U limit (U/t=10), and is concomitant with a renormalization of the electronic mass (see fig.S1.c). We associate hence the CDW to a charge ordering induced by the narrowing of the conduction band induced by correlation. The electronic localization is associated with the opening of a pseudo-gap.

\section{Acknowledgements}

This research was supported by EPSRC (EP/M011038/1 and also EP/R02992X/1) and the Simons Many-Electron Collaboration. C.W. gratefully acknowledges the support of NVIDIA Corporation, ARCHER UK National Supercomputing Service. We are grateful to the UK Materials and Molecular Modelling Hub for computational resources, which is partially funded by EPSRC (EP/P020194/1).


\section{Method}


In this paper we have used the following computational setup in order to model the deposition of 
Ce adatoms on Ag substrate: the Ag substrate was modelled by a 2D triangular lattice of
Ag 5$s$ orbitals, while the Ce adatom 4f orbital was represented by an additional site connected
with its three nearest neighbor Ag atoms.
The heavy deposited Ce adatom hybridize with the substrate, and also induce an attractive ionic potential on the three neighbors of the substrate. 

In this work, we use the dynamical mean-field theory (DMFT) approximation.
Heavy f-materials have been extensively studied by DMFT, which captures well both the itinerant and localized behavior of the electrons in f-materials~\cite{Pourovskii_2007,Amadon_2008,Lichtenstein2013,Lechermann_2006,Plekhanov2018}.
In the DMFT formalism, the adatom (Ce) sites with the on-site $U$ are treated as strongly correlated orbitals, while the conduction band of the substrate (Ag sites) are considered as uncorrelated. More formally, the Hamiltonian of the system
considered here reads in real-space as:
\begin{equation}
\begin{split}
H= & -t\sum_{\left\langle i,j\right\rangle
,\sigma}c_{i\sigma}^{\dagger}c_{j\sigma}+\sum_{\sigma,k}\epsilon_{\text{imp}}f_{\sigma,k}^{\dagger}
f_{\sigma,k}+\sum_{i,\sigma}V_{i}c_{i\sigma}^{\dagger}c_{i\sigma}\\
& +\sum_{\left\langle i,k\right\rangle
,\sigma}(V_{if}c_{i\sigma}^{\dagger}f_{\sigma,k}+h.c)+U\sum_{\sigma,k}{n_{f\uparrow,k}n_{f\downarrow,k}}+\mu\hat{N}_{tot},
\end{split}
\label{AIM}
\end{equation}
where $t$ is the
inter-substrate hopping (in our calculation $t$ is the energy scale, for the case of a Ag(111) substrate, $t=0.75$eV), $\epsilon_{imp}$ is the adatom energy level, $U$ the adatom Hubbard repulsion
and $\mu$ the chemical potential. $c^{(\dagger)}_{i\sigma}$ and $f^{(\dagger)}_{i\sigma}$ are the Fermionic 
annihilation (creation) operators of the site $i$, spin $\sigma$, surface state and the adatom
respectively. $V_i$ describes the on-site potential asserted upon the substrate by the adatom (as
described in \citep{cedric2004}). And lastly, the adatom/substrate hopping parameter
$V_{if}=V^*_{fi}=t_{Ad}$ is assumed finite for three of the substrate sites, otherwise it is zero.
This assumption can be justified through consideration of the triangular
substrate\cite{blackbourne,cedric2004}. The schematic view of our computational setup is shown in
fig.\ref{scheme}.


In analogy with DFT+DMFT, the calculation requires convergence with respect to the number of
$k$-points in the Brillouin zone of substrate lattice.
Calculations carried out in this work were computationally intensive, convergence in some cases
required $100\times100$ sixfold reduced $k$-points (triangular substrate symmetry). Without any
simplifications, calculations of each $k$-point scales as $n_{freq}N^3$, for number of sites $N$ and
number of Green's function frequency points $n_{freq}=\mathcal{O}(1000)$. For the largest lattice
size considered, $L=15$, of the order of $1000*225^3$ calculations was needed to invert the Hamiltonian to
calculate the Green's function for a single DMFT iteration.

In DMFT, we have employed an exact diagonalization (ED) solver improved by using the cluster
perturbation theory (CPT). The ED solver\cite{ed1,ed2,ed3} uses a finite  discretization approximation of 
hybridization $\Delta(i\omega_n)$ function, which induces finite discretization errors \cite{lanczos}. As the
size of the basis used to represent the Hamiltonian scales exponentially with the number of bath energies in the system, solving the eigenvalue problem is not computationally achievable if the number of bath orbitals is typically larger than 15.
ED-CPT\cite{cedric2012} bypasses this problem by utilizing a small number of bath systems and
incorporating additional bath sites in order to mimic the effect of coupling to an infinite bath, which was used in our work to check that no remaining systematic error due to the bath discretization was obtained in our calculations.
Finally, we have used Lichtenstein's double counting scheme~\cite{LichtensteinDC}.

\subsection{Sherman-Morrison}

Calculation of the full lattice Green's function $G(k)$ involves a matrix inversion at each
$k$-point and $\omega$. In the case of large supercell sizes this becomes a bottleneck of the formalism. An optimization is possible here due to the fact that the matrix to be inverted
 contains a large number of empty
entries. This is due to the properties of the supercell: the self energy is
non-zero only at the adatom, and the hybridization between the adatom and the 
substrate occurs only on three sites of the substrate. Therefore, to reduce the 
computational load, the Sherman-Morrison algorithm \cite{sherman} was
implemented to update the full Green's function $G(k)$ from the substrate
Green's function $G_0(k)$:

\begin{equation} 
   G_0^{-1}(k) =
   \begin{bmatrix}[ccccc] 
	  i\omega-V_i & -t & -t &  0 & \dots \\ 
	  -t & i\omega-V_i & -t & -t & \dots \\
	  -t & -t & i\omega-V_i & -t & \dots \\ 
	   0 & -t & -t & i\omega     & \dots \\
	   \vdots & \vdots & \vdots & \vdots & \ddots
   \end{bmatrix},
\end{equation}

\begin{equation} 
\setlength{\arraycolsep}{2pt}
\renewcommand{\arraystretch}{1}
   G^{-1}(k) =
   \begin{bmatrix}[ccccc|c] 
	   & & & & & V_{if} \\
	   & & & & & V_{if} \\
	   & & G_0^{-1}(k) & & & V_{if} \\ 
	   & & & & & 0 \\ 
	   & & & & & \vdots\\ 
	  \hline V_{if} & V_{if} & V_{if} & 0 & \dots & \epsilon_f-\Sigma \\
   \end{bmatrix}.
\end{equation}

Here $V_{if}$, $\epsilon_f$, $t$, and $V_i$ are defined in~(\ref{AIM}), $i \omega_n$ is the $n$'s Matsubara frequency, while the substrate sites are ordered in such a way that the three sites connected to the adatom are listed first.

$G_0$ is obtained via matrix products for all $n_{freq}$ ($N^2$
operations), $G$ is then updated from $G_0$. Since only three substrate
sites hybridizes with the adatom, the updating process scales linearly
with $N$.

\subsection{Kondo physics}

In order to extract the Kondo physics fingerprints from our DMFT results we have calculated 
the scattering rate $\Gamma$ and Kondo temperature $T_k$.

In order to estimate $\Gamma$, we performed the polynomial regression with the following low
frequency, low temperature expansion of $\Sigma''(i\omega_n)$ in the Kondo regime\cite{costi}:

\begin{equation}
\Sigma''(i\omega_n) = a + \left(1-\frac{1}{Z}\right)\omega_n - \frac{1}{2Z^2\Delta''(i\omega_n)} \omega_n^2.
\end{equation}

$\Gamma$ is obtained from the above parameters $a$ and $Z$ as follows:

\begin{equation}
\Gamma=-\frac{a}{Z}.
\end{equation}

We used \cite{werner} the following definition of the Kondo temperature $T_k$ after Hewson's
renormalized perturbation theory\cite{hewson} of the Anderson impurity model (AIM):
\begin{equation}
T_k=-\frac{\pi Z}{4}\Delta''(i\omega_n)\mid_{\omega_n\rightarrow0}
\end{equation}
where $\Delta''$ is the imaginary part of the resonant level scattering width as defined in Ref.\citep{hewson}.

 \subsection{Triplet bound state toy model}

Here we briefly outline the simple toy model used to analyze the regime of the phase diagram, where we have two electrons per unit-cell. As we observed in the superlattice calculations that the electronic states are localized around the adatoms, we solve the isolated plaquette of three substrate atoms connected to the adatom site. This Hamiltonian includes a local Coulomb repulsion $U$ on the adatom site, an hybridization between the adatom and the three substrate atoms, and hopping across the substrate atoms. The impurity energy and hybridization are the same as the one treated in the full DMFT calculations, whereas the hopping across the substrate elements is renormalized to account for the bandwidth of the full calculation. Upon diagonalizing this Hamiltonian and finding the ground state, we then evaluate the expectation value of the spin correlator $K$ operator between the adatom and one of the substrate atom, which identifies a singlet (triplet) ground state if $K<0$ ($K>0$). We identify the first order transition between the triplet state at small $U$ and the singlet state at large $U$ (see fig.S6). Interestingly, we find that the critical $U$ to obtain a singlet state is larger than all realistic $U$ considered in our calculation. This suggests that in the limit of low occupation, pairs of electrons will form a triplet state for realistic values of the Coulomb repulsion.

This triplet state is indeed a multiplet that also includes the polarized contribution with two electrons of same spin, and exhibit a lower degree of correlation. Indeed, in the $S_{tot}^{z}=\pm 1$ sectors, there are no contributions from the Coulomb energy, as observed in the phase diagram near the $n_f=0.5$ region, where the mass renormalization is weak. We attribute hence this region of the phase diagram to a triplet bound state.








\bibliography{bibliography} 

\onecolumngrid
\newpage

\section{\huge{\centerline{Supplementary Material}}}
\setcounter{page}{1}


\begin{figure}[h!]
\includegraphics[width = 1.075\textwidth, trim= 1.0cm 1.0cm 1cm -18cm]{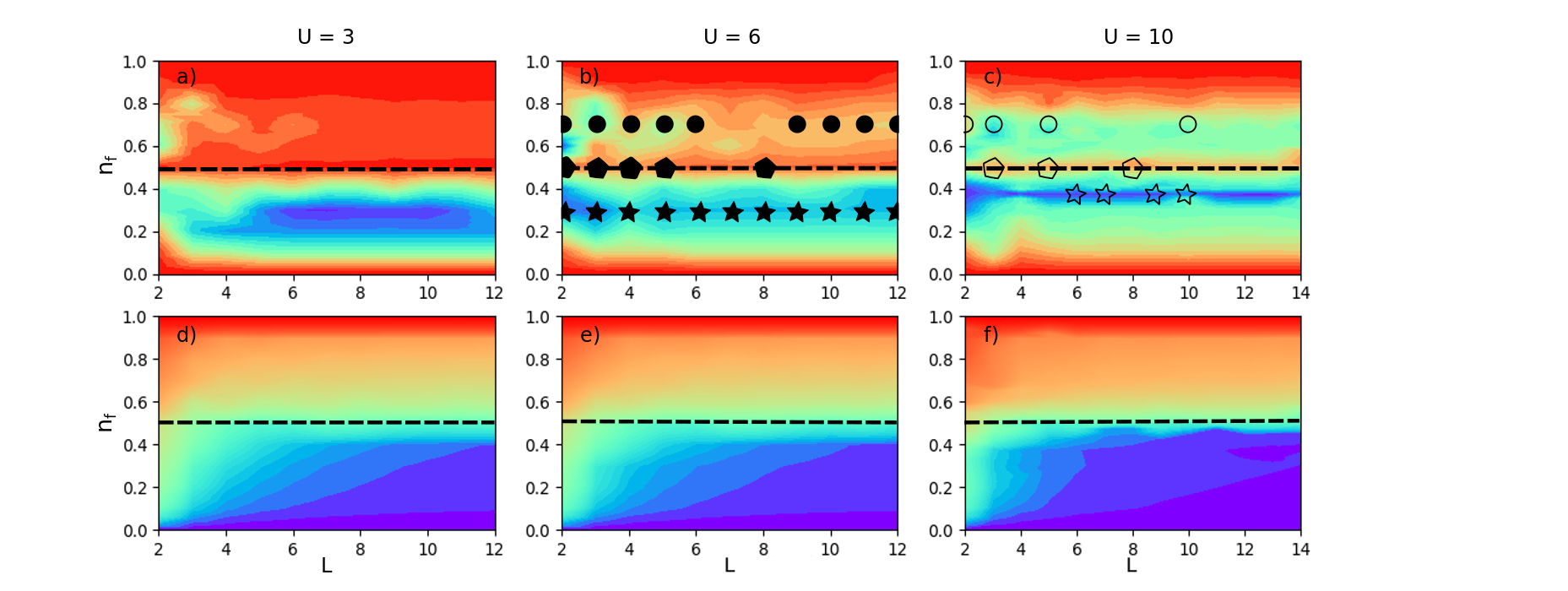}
\caption{(colour online) Upper panel: colour map of the quasi-particle weight $Z$ for a) $U=3$, b) $U=6$ and c) $U=10$ for different lattice sizes $L$. $Z$ is a measure of the strength of the electronic correlations induced by $U$, where the colour map indicates $Z=0$ (blue) and $Z=1$ (red). We identified three regions of interest in the phase diagram: for $n_f=0.3$, $n_f=0.5$ and $n_f=0.7$. We report further in the text observables that characterize these three regions, which are identified by the filled and open symbols.  Lower panel: colour map of the substrate (conduction band) electronic density for different substrate sizes $L$ for d) $U=3$, e) $U=6$ and f) $U=10$ (blue for $n_{CB}=0$ and red for $n_{CB}=1$). The conduction band's electronic states are empty in the lower right corner of the phase diagram, which indicates that in this region the adatom is decoupled from the substrate.}
\label{atlas2}
\end{figure}
\vspace{-17.cm}

\begin{figure}
\includegraphics[width = 1\columnwidth, trim= 0cm 10.5cm 0cm 11cm]{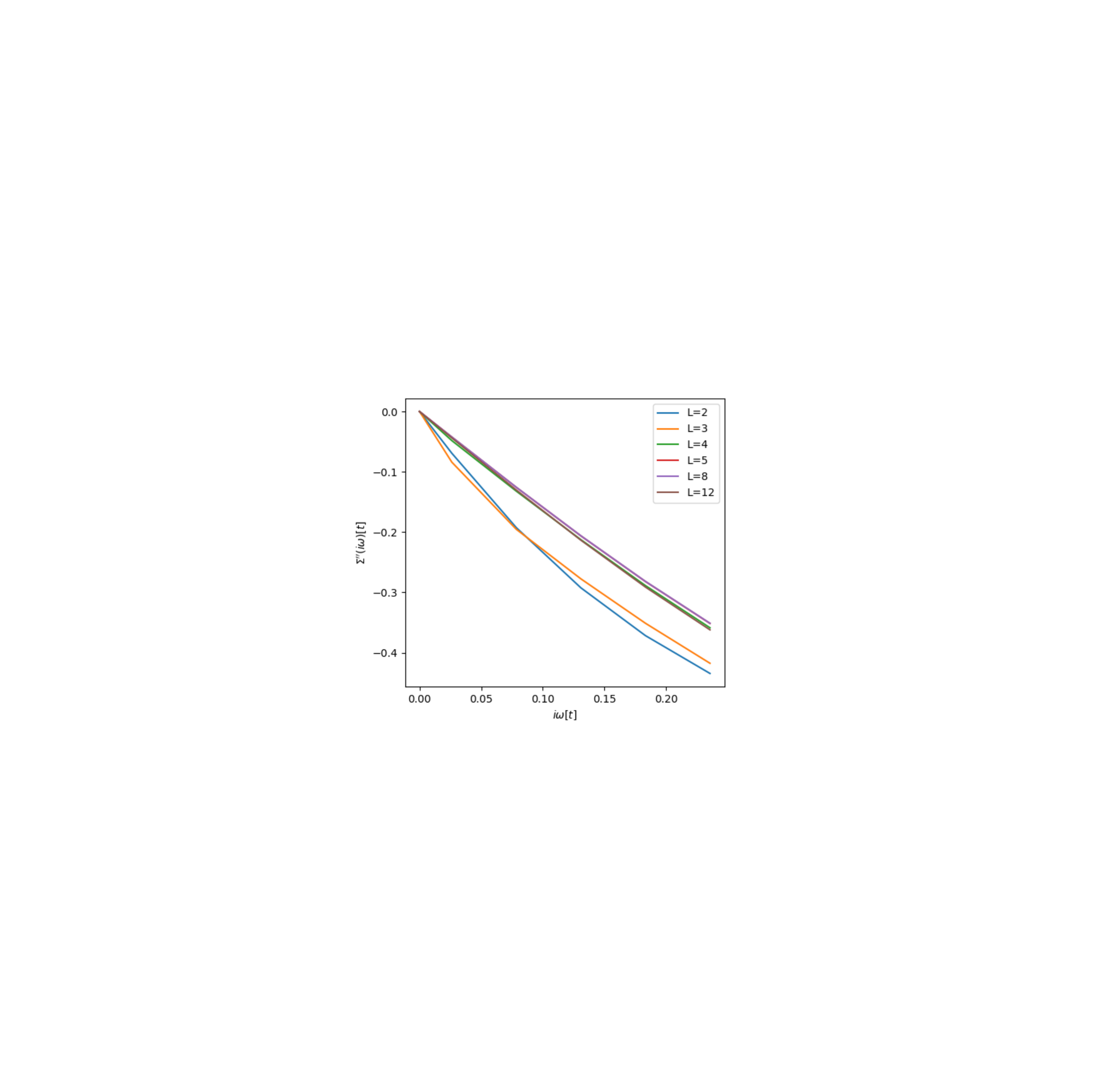}
\includegraphics[width=0.5\columnwidth]{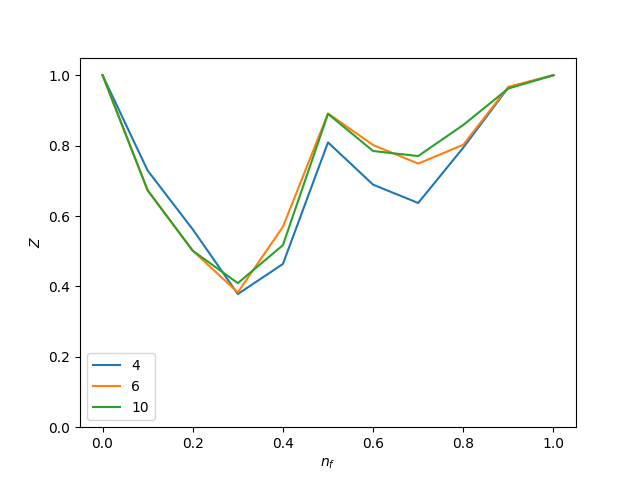}
\caption{(colour online) (Upper panel) Imaginary part of adatom self-energy $\Sigma''$ as a function of Matsubara frequency $i\omega$, at $U=6$ and $n_f=0.3$ across a range of $L$.  (Lower panel) Adatom quasi-particle weight $Z$ as a function of $n_f$ across a range of L. We observe two regions of correlated electron physics centered at $n_f=0.3$ and $n_f=0.7$. }
\label{sigma''}
\end{figure}

\begin{figure*}[h!]
\includegraphics[width = 1\textwidth, trim= 13.0cm 12.0cm 13cm 12.0cm]{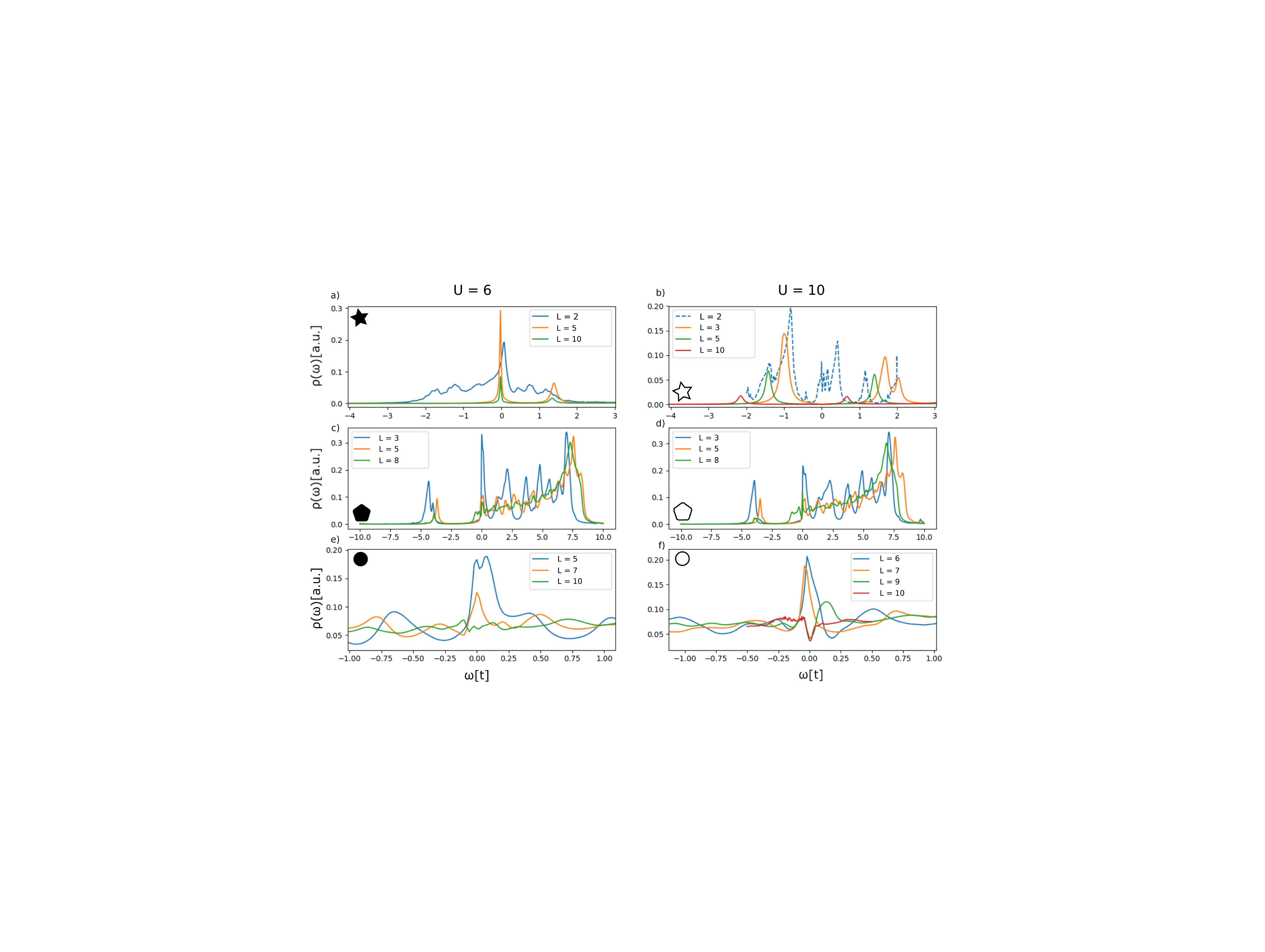}
\caption{(colour online) Normalized density of states $\rho(\omega)$ of the supercell for $U=6$ (left column) and $U=10$ (right column) obtained at the points highlighted in fig.\ref{atlas2}. For low adatom electron densities $n_{f}\sim0.3$ correlation effects are observed at low lattice sizes $L$ which split into well-defined peaks at larger $L$. For high $n_{f}$, Kondo peaks are observed at low $L$, which disappears/ forms dips at high $L$.}
\label{dos1}
\end{figure*}
\begin{figure*}[h!]
\includegraphics[width = 0.45\columnwidth, trim = 13cm 12cm 12cm 12cm]{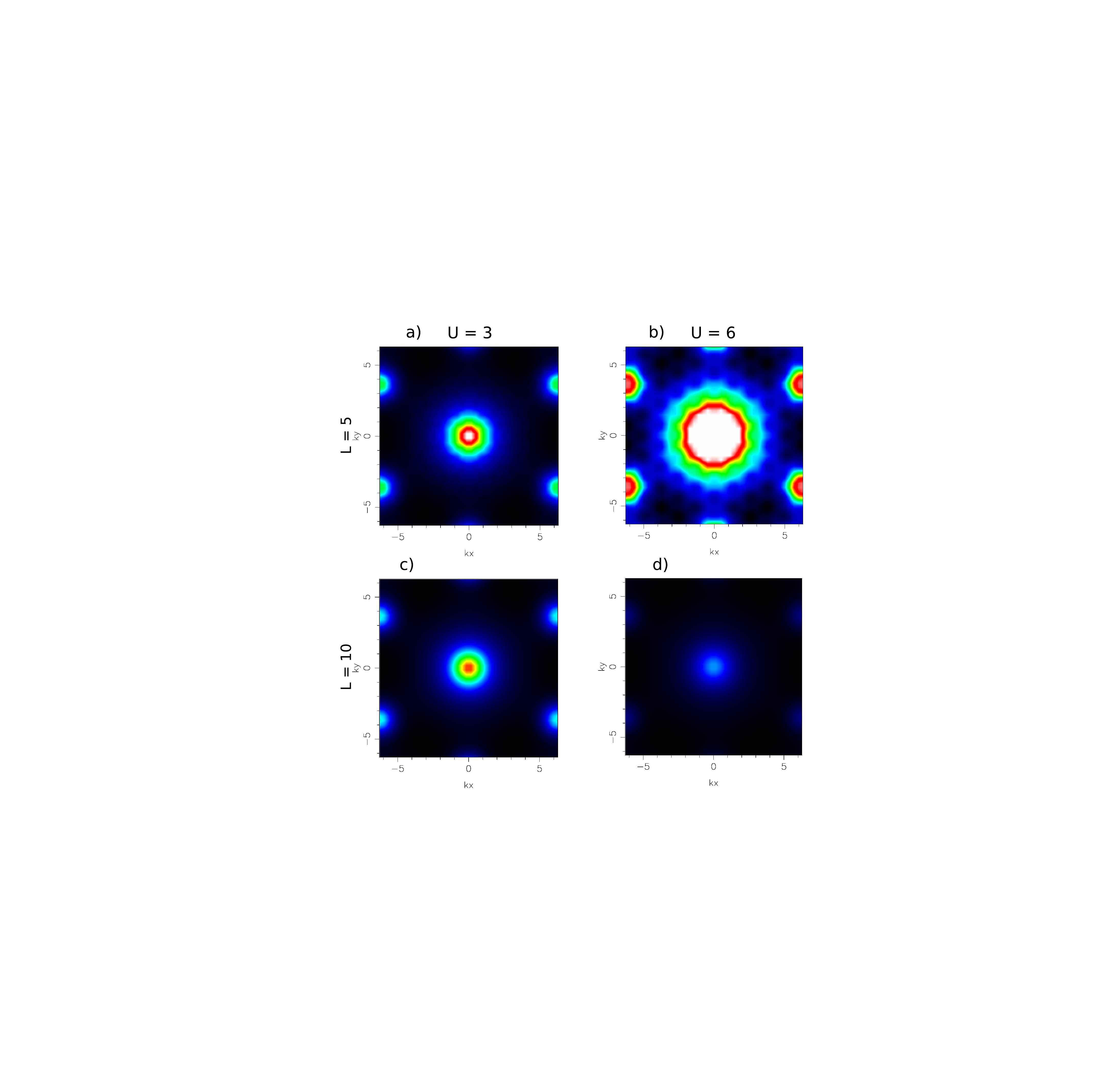}
\caption{(colour online) Unfolded Fermi surface colour map at adatom electron density $n_{f}\sim0.3$ for $U=3,6$ and lattice size $L=5,10$ at Fermi energy $\epsilon_f$. We observe a decrease in electronic momentum for $U=6$ as $L$ increases, this indicates a decoupling between the adatom and the substrate.}
\label{FERMI}
\end{figure*}


\begin{figure*}[h!]
\includegraphics[width=0.5\columnwidth]{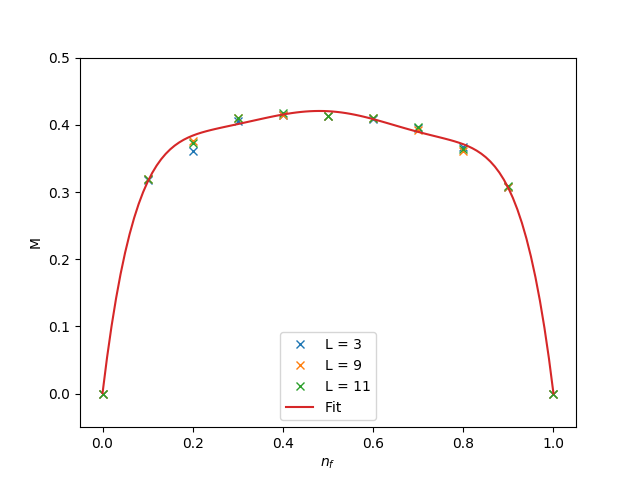}
\caption{(colour online) Local magnetic moment $M$ of the adatom f state as a function of the adatom electronic occupation. A multivariate fit is shown as a guide to the eye. $M$ is weakly dependent on the geometry of the conduction band and shows that the magnetic moment is well captured at the level of a local approximation.}
\label{stot}
\end{figure*}

\begin{figure}[h!]
\includegraphics[width=0.45\columnwidth]{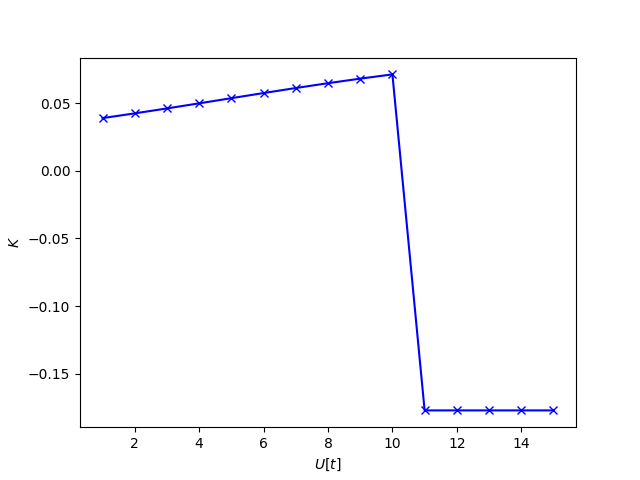}
\caption{Spin correlator $K$ of the adatom and substrate atoms. At hopping $t=1$, adatom energy $\epsilon_{imp}=-6$ and adatom-substrate hybridzation $V_{if}=3$ across a range of value of $U$. We observe a triplet ($U<10$) to singlet transition ($U>11$).}
\end{figure}

\end{document}